\documentclass[aps,prd,twocolumn,showpacs,groupedaddress]{revtex4}
\usepackage{graphicx}
\usepackage{dcolumn}
\usepackage{bm}
\usepackage{color}
\usepackage{longtable}
\usepackage{supertabular}
\usepackage[T1]{fontenc}

\makeatletter

\newcommand{\Rmnum}[1]{\expandafter\@slowromancap\romannumeral #1@}
\makeatother

\begin{document}
\title{Search for Lorentz invariance violation through tests of the gravitational inverse square law at short-ranges}
\author{Cheng-Gang Shao$^{1}$}
\author{Yu-Jie Tan$^{1}$}
\author{Wen-Hai Tan$^{1}$}
\author{Shan-Qing Yang$^{1}$}
\author{Jun Luo$^{1}$}
\email[E-mail: ]{junluo@sysu.edu.cn}
\author{Michael Edmund Tobar$^{2}$}
\email[E-mail: ]{michael.tobar@uwa.edu.au}

\affiliation{
$^{1}$Department of Physics, Huazhong University of Science and Technology, Wuhan 430074, People's Republic of China\\
$^{2}$School of Physics, University of Western Australia, Crawley, WA 6009, Australia
}

\date{\today}

\begin{abstract}
A search for sidereal variations in the non-Newtonian force between two tungsten plates separated at millimeter ranges sets experimental limits on Lorentz invariance violation involving quadratic couplings of Riemann curvature. We show that the Lorentz invariance violation force between two finite flat plates is dominated by the edge effects, which includes a suppression effect leading to lower limits than previous rough estimates. From this search, we determine the current best constraints of the Lorentz invariance violating coefficients at a level of $10^{-8}$ m$^{2}$.
\end{abstract}


\maketitle
Local Lorentz invariance is at the foundation of both the Standard Model of particles physics and General Relativity (GR), however, the later theory is formulated as a classical theory, which demands some changes in its foundational structure to merge gravity with quantum mechanics. Even if local Lorentz invariance is exact in the underlying theory of quantum gravity, spontaneous breaking of this symmetry may occur, leading to tiny observable effects \cite{ksp1,ksp2}. On the other hand, Lorentz violations could also be large but "hard-to-see", despite many experiments to date setting very tight bounds across many physical sectors \cite{tables}. This would occur if the Lorentz invariance violation is "countershaded" as pointed out in ref \cite{kt}. Thus in general, the investigation of local Lorentz invariance violations in the spacetime theory of gravity is a valuable tool to probe the foundations of GR\cite{1,2} without preconceived notions of the numeric sensitivity.

Recently, the studies of Lorentz invariance violation in the pure-gravity sector shows that general quadratic curvature coupling will lead to interesting new effects in short-range experiments that could have escaped detection in conventional studies to date \cite{3}. Accordingly a crude estimation has been made on the possible constraints for these types of Lorentz invariance violation\cite{3}, predicted to be tested to the level of $10^{-8}$ to $10^{-10}$ m$^{2}$ using short-range experiments, such as the EotWash \cite{4,5}, Wuhan \cite{6,7}, and Bloomington\cite{8} experiments. However, for the latter two results, edge effects are not considered properly. In this work we obtain the best current constraint of the Lorentz invariance violation at level $10^{-8}$ m$^{2}$ from a detailed reanalysis of prior data taken from the Wuhan experiment (HUST-2011).

Effective field theory is a powerful and unique tool for investigating physics at attainable scales, and is suited for exploration of local Lorentz invariance in gravity. For centuries after Newton's Principia, our experimental understanding of gravity remains in some respects remarkably limited. We are confident that Newton's law describes the dominant physics in long-range gravity and GR provides accurate relativistic corrections. However, in short-range gravity, it is presently unknown whether gravity obeys Newton's law, and many models attempting to unify gravity and the other fundamental forces in the same theoretical framework predict modifications of gravity, i.e., a deviation of the Newtonian $1/{r^{2}}$ law. The experiments in short-range gravity are well suited for exploration of local Lorentz invariance for spacetime-based gravitation. Here we use the data from a shot-range experiment in HUST-2011 to perform a search for short-range Lorentz invariance violation involving quadratic couplings of Riemann curvature.

A quantitative description of Lorentz invariance violation in the pure-gravity sector with quadratic couplings of Riemann curvature is given by the Standard-Model Extension (SME) \cite{tables}. Using effective field techniques, the coupling leads to perturbative corrections to Newton gravity, which are inverse quartic and vary with orientation and time. The corresponding perturbative corrections between two test masses $m_{1}$ and $m_{2}$ in the general quadratic curvature coupling of the SME is given by \cite{3}\begin{eqnarray}\label{equation1:eps}
{V_{LV}}(\vec{x})
=\!\!\!\!\!\!&&- G \frac{{m_1}{m_2}}{|\vec{x}_{1}-\vec{x}_{2}|^{3}}
\Bigl[\frac{3}{2}{({{\bar k}_{\rm{eff}}})_{jkjk}}-9{({{\bar k}_{\rm{eff}}})_{jkll}}\hat{x}^{j}\hat{x}^{k}\nonumber\\*
&&+\frac{15}{2}{({{\bar k}_{\rm{eff}}})_{jklm}}
\hat{x}^{j}\hat{x}^{k}\hat{x}^{l}\hat{x}^{m}
\Bigr]\\*
\equiv \!\!\!\!\!\! && - G \frac{{m_1}{m_2}}{|\vec{x}|^{3}}{\bar k}(\hat{x},T),\nonumber
\end{eqnarray}
where $\vec{x}=\vec{x}_{1}-\vec{x}_{2}$ is the vector separating $m_1$ and $m_2$, $\hat{x}^{j}$ is the projection of the unit vector along $\vec{x}$ in the $j$th direction. $({{\bar k}_{\rm{eff}}})_{jklm}$ is a set of 81 coefficients for Lorentz invariance violation in standard laboratory, which is totally symmetric with indices $j$, $k$, $l$, $m$ ranging over the three spatial direction, representing 15 independent observables for Lorentz invariance violation. These coefficients can be taken as constant on the scale of the solar system. Therefore the canonical frame adopted for reporting results from experimental searches for Lorentz invariance violation is the Sun-centered frame\cite{sunframe1,sunframe2,sunframe3}, with $Z$ axis along direction of the Earth's rotation and $X$ axis pointing towards the vernal equinox.  The zero in time is taken to be the vernal equinox at March 20, 2000 at 7:35 a.m. Universal Time. Neglecting the Earth's boost, which is older $10^{-4}$, the transformation from the Sun-centered frame $(X, Y, Z)$ to the laboratory frame $(x, y, z)$ involves a time-dependent rotation $R^{jJ}$. Taking the laboratory $z$ axis pointing to the local zenith and the angle between $x$ axis and local south being $\theta$, the rotation matrix is
\begin{equation}\label{equation2:eps}
\begin{array}{c}
{R^{jJ}} = \left[ {\begin{array}{*{20}{c}}
\cos\theta & \sin\theta & 0\\
-\sin\theta & \cos\theta & 0\\
0& 0 & 1
\end{array}} \right]\\
\qquad
\left[ {\begin{array}{*{20}{c}}
\cos\chi\cos\omega_{\oplus}T & \cos\chi\sin\omega_{\oplus}T &-\sin\chi\\
-\sin\omega_{\oplus}T & \cos\omega_{\oplus}T & 0\\
\sin\chi\cos\omega_{\oplus}T & \sin\chi\sin\omega_{\oplus}T & \sin\chi
\end{array}} \right],
\end{array}
\end{equation}
where $\chi$ is the colatitude angle of the laboratory and $\omega_{\oplus}\simeq 2\pi/$(23h56min) is the Earth's sidereal frequency. The angle $\theta$ is constant depending on the orientation of the apparatus in the experiment (For the experiment in Wuhan, we take $\theta=-\pi/2$.). The sidereal time $T$-dependent coefficients $({{\bar k}_{\rm{eff}}})_{jklm}$ are thus related to the constant coefficients $({{\bar k}_{\rm{eff}}})_{JKLM}$ in the Sun-centered frame by
\begin{equation}\label{equation3:eps}
({{\bar k}_{\rm{eff}}})_{jklm}=R^{jJ}R^{kK}R^{lL}R^{mM}({{\bar k}_{\rm{eff}}})_{JKLM}
\end{equation}
Therefore, the inverse-cube potential for Lorentz invariance violation in Eq.(1) is oscillatory with $T$ and includes components up to the fourth harmonic of $\omega_{\oplus}$, which leads to striking signals in short-range experiments. Equivalently, the Lorentz invariance violation force between two plates can be expected to vary with frequencies up to and including the fourth harmonic of $\omega_{\oplus}$.

Most Inverse-Square Law tests use planar geometry to search for the non-Newton gravity. Noting that the Newtonian force at any point above an infinite plane of uniform mass density is constant, the planar geometry can be effective in suppressing the Newtonian background relative to the putative short-range effect. However, it also suppresses the Lorentz invariance violation signal produced by Eq. (1). For instance, the Lorentz invariance violation force between a flat plate and an infinite plane is zero (shown in Fig. 1) due to the fact that the sum of the three coefficients in Eq.(1) is zero. The shape and edge effects play an important role in determining the sensitivity of the experiment to the coefficients for Lorentz invariance violation and is explained in following.

Let's consider the Lorentz invariance violation interaction in Eq.(1) between a flat plate of area $A_{P}$, thickness $t_{P}$, and density $\rho_{P}$ at distance $d$ from an infinite plane of thickness $t$ and density $\rho$. Assuming the normal vector of the plates is in $y$ axis direction as shown in Fig. 1, the Lorentz invariance violation acceleration on a point above the infinite plate with distance $d$ is
\begin{equation}\label{equation4:eps}
a_{LV}^y(d) = G\rho \left[ {1 - \frac{d}{{d + t}}} \right]\int_0^{2\pi } {d\varphi_{p} } \int_0^\infty  {rdr} \frac{{\bar k(\theta_{p} ,\varphi_{p} )}}{{{{\sqrt {{r^2} + {d^2}} }^3}}},
\end{equation}
where the integration of the direction dependence $\bar k(\theta_{p},\varphi_{p} )$ can be carried out over the entire volume of the infinite plate. The integral in Eq(\ref{equation4:eps}), shows that the Lorentz invariance violation force between the plates is
\begin{eqnarray}\label{equation5:eps}
F_{LV}^y(d)\mid_{infinite}=\!\!\!\!&&C\Bigl[\frac{3}{2}{({{\bar k}_{\rm{eff}}})_{jkjk}}-9{({{\bar k}_{\rm{eff}}})_{jkll}}\frac{\delta_{jk}}{3}\nonumber\\*
&&+\frac{15}{2}{({{\bar k}_{\rm{eff}}})_{jklm}}\frac{{\delta_{jk}}{\delta_{lm}}}{5}\Bigr]=0
\end{eqnarray}
with $C \equiv 2\pi G{\rho _p}\rho {A_p}\ln \frac{{(d + {t_p})(d + t)}}{{(d + {t_p} + t)d}}$.

\begin{figure}[hp]
\includegraphics[width=0.4\textwidth]{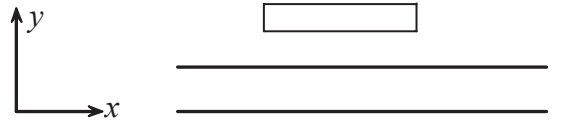}
\caption{\label{fig:fig_1} The inverse-cube potential between two plates. Lorentz invariance violation force between a flat plate of area $A_{P}$, thickness $t_{P}$, and density $\rho_{P}$ at distance $d$ from an infinite plane of thickness $t$ and density $\rho$ is zero, which means the Lorentz invariance violation force between two finite flat plates is dominated by the edge effects.}
\end{figure}

For the finite plates used in most experiments, one must use numerical integration. Noting that the measured signal in experiments is usually the variation of the force when the gap between two plates changes from $d_{\rm{min}}$ to $d_{\rm{max}}$, the variation of the Lorentz invariance violation signal can be roughly estimated as
\begin{equation}\label{equation6:eps}
\Delta F_{LV}^y = F_{LV}^y({d_{\rm{\min }}}) - F_{LV}^y({d_{\rm{\max }}})\sim\varepsilon \Delta C{({\bar k_{\rm{eff}}})_{jklm}}
\end{equation}
with
\begin{eqnarray}\label{equation7:eps}
\Delta C \equiv \!\!\!\!\! && 2\pi G{\rho _p}\rho {A_p}\Bigl[ \ln \frac{({d_{\min }} + {t_p})({d_{\min }} + t)}{({d_{\min }} + {t_p} + t){d_{\min }}} \nonumber\\*
&&- \ln \frac{({d_{\max }} + {t_p})({d_{\max }} + t)}{({d_{\max }} + {t_p} + t){d_{\max }}} \Bigr]
\end{eqnarray}
and the dimensionless parameter $\varepsilon$ represents edge effects, typically of order $10^{-2}$ or $\sim$d/$\sqrt{A}$ in most inverse square law experiments.
\begin{figure}[!t]
\includegraphics[width=0.50\textwidth]{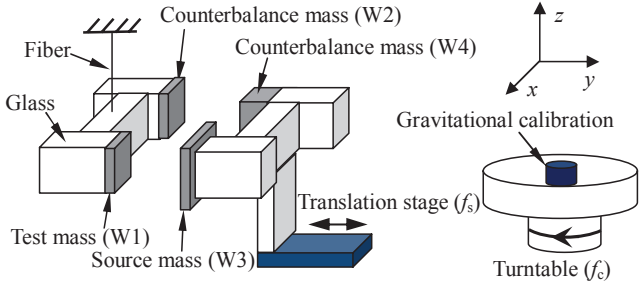}
\caption{\label{fig:fig_2} Schematic drawing of the experiment HUST-2011. The pendulum twist was measured by an autocollimator, and controlled by applying differential voltages to the two capacitive actuators (not shown here). The separation between the counterbalance masses W2 and W4 ($\approx$ 4mm) is larger than that between the test mass W1 and source mass W3. The separation between W1 and W3 was modulated by driving a motor translation stage with frequency $f_{\rm{s}}$, and the sensitivity of the pendulum was calibrated by rotating the copper cylinder at frequency $f_{\rm{c}}$.}
\end{figure}

We then consider the experiment for HUST-2011. The design and operation of the experiment are described in Ref. \cite{7}. Here we summarize briefly the basic features as shown in Fig.2. Two pure tungsten plates with the same dimensions of 16mm$\times$16mm$\times$1.8mm used as the test mass (W1) and the counterbalance one (W2), were adhered to the two sides of the I-shaped pendulum symmetrically. The arm length of the test mass is 42mm. The source mass platform, facing to the pendulum, was also designed as I-shaped structure. A larger tungsten source mass (W3) with dimension 20.8mm$\times$20.8mm$\times$1.8mm was adhered to the one side of the I-shaped structure, opposite to the test mass W1. The non-Newtonian force to be measured is between W1 and W3 for separations ranging from 0.4 to 1.0 mm, where the net torque change of the Newtonian force was counteracted by a thicker tungsten counterweight mass (W4) with dimensions of 16mm$\times$16mm$\times$7.6mm, adhered on the other side of the source mass platform, located behind W3 about 4 mm in $y$ axis direction. The normal vectors of the plates are all in $y$ axis direction in local laboratory frame (Fig.1 and Fig.2). Taking $A_{P}$=16mm$\times$16mm, $t_{P}$=$t$=1.8mm, arm length $l$=42mm, $d_{\rm{min}}$ =0.4mm, and $d_{\rm{max}}$ =1.0mm, we obtain $\Delta {C_{{\rm{HUST}}}}l \approx 9.42 \times {10^{ - 7}}{\rm{Nm/}}{{\rm{m}}^{\rm{2}}}$ according to Eq.(7). The numerical integration gives the torque for Lorentz invariance violation between test mass and source mass, and it is determined to be
\begin{eqnarray}\label{equation8:eps}
\Delta \tau _{LV}^z \approx \!\!\!\!\!\! &&\left[ { - 0.06{{({{\bar k}_{\rm{eff}}})}_{xxxx}} + 0.16{{({{\bar k}_{\rm{eff}}})}_{xxyy}} +  \cdots } \right] \nonumber\\*
&&\times {10^{ - 7}}{\rm{Nm/}}{{\rm{m}}^{\rm{2}}} <  < \Delta {C_{{\rm{HUST}}}}l{\bar k_{\rm{eff}}},
\end{eqnarray}
which means the edge effect suppression factor is approximately $\varepsilon\sim$0.01. Taking the sensitivity of the torque as 0.8$\times$10$^{-16}$ Nm, the constraint of $(\bar{k}_{\rm{eff}})$ can be roughly estimated at level of 0.5$\times$10$^{-8}$ m$^{2}$. Then the detailed analysis of sidereal variations of torque can set the experimental limits on Lorentz invariance violation parameters, $(\bar{k}_{\rm{eff}})_{JKLM}$, in the Sun-centered frame.

In the experiment for HUST-2011, the motion of the pendulum was controlled by using a proportional-integral-differential feedback system. The separation between the tungsten test and source masses was modulated by driving a motor translation stage with frequency $f_{\rm{s}}=2\pi/500$s, and the sensitivity of the pendulum was calibrated by rotating a copper cylinder at frequency $f_{\rm{c}}=2\pi/400$s. The recorded data involves 12 sets, acquired from August to October in 2011. Each set has a duration time of approximately one day. To extract the expected signal, each data run was broken into separate ``cuts'' containing exactly the lowest common multiple of the dual-modulation periods, $\Delta T$ =2000s. For each ``cut'', and then the non-Newton torque was fitted with
\begin{equation}\label{equation9:eps}
\tau _{measured}^z(T) = {\tau_{LV}}(T)\cos (2\pi {f_{\rm{s}}}T + \varphi )
\end{equation}
with the initial phase $\varphi$ for the separation modulation, which can be determined according to the operation procedure in the experiment. For averaging times of  $\Delta T$ =2000s, the sidereal variation of ${\tau_{LV}}(T)$ can be neglected since ${\omega _ \oplus }\Delta T <  < 1$. Thus, we take ${\tau_{LV}}(T)$ as constant approximately in each ``cut''. Fig.3 represents the sidereal variation of ${\tau_{LV}}(t)$ varying with time $t$. Each point represents the mean of a 2000 second measurement without showing the error bar, which is dominated by statistical uncertainty. In the Sun-centered frame, the time scale is reset by $t=0$ when the Sun passes the vernal equinox on 2011, March 20, 23 h 21 min universal time (UT). In the laboratory frame, the time scale is set by $T=0$ when the east direction coincides with the Y axis. The difference between the two time scales can be written as a phase difference $\omega _ \oplus(T-t)\simeq-1.35$ for this experiment (east longitude 114.4$^\circ$ and $\chi=59.5^\circ$).

According to Eqs. (2) and (3), ${\tau_{LV}}(T)$ can be further expressed as a Fourier series in the sidereal time $T$
\begin{eqnarray}\label{equation10:eps}
{\tau _{LV}}(T) = \!\!\!\!\!\! &&{c_0} + \sum\limits_{m = 1}^4 \left[ {c_m}\cos (m{\omega _ \oplus }T) + {s_m}\sin (m{\omega _ \oplus }T) \right]\nonumber\\*
&&\frac{\sin (m{\omega _ \oplus }\Delta T/2)}{m{\omega _ \oplus }\Delta T/2},
\end{eqnarray}
in which we have considered corrections due to data averaging. For the sinusoidal signal with angular frequency $m{\omega _ \oplus }$, the data average over  $\Delta T$ intervals will lead to the attenuation of amplitude by a small amount $1 - \frac{{\sin (m{\omega _ \oplus }\Delta T/2)}}{{m{\omega _ \oplus }\Delta T/2}}$, which is approximately zero. The Fourier amplitudes in this expression are functions of the coefficients $(\bar{k}_{\rm{eff}})_{JKLM}$, the test mass and source mass geometry, and the laboratory colatitude. For example, the coefficients $c_{0}$ and $c_{1}$ related to $(\bar{k}_{\rm{eff}})_{JKLM}$ can be obtained by numerical integration,
\begin{equation}\label{equation11:eps}
\begin{array}{c}
{c_0} = [ - 0.012{({{\bar k}_{\rm{eff}}})_{XXXX}} - 0.012{({{\bar k}_{\rm{eff}}})_{YYYY}}\\
 + 0.025{({{\bar k}_{\rm{eff}}})_{ZZZZ}} - 0.024{({{\bar k}_{\rm{eff}}})_{XXYY}}\\
 + 0.011{({{\bar k}_{\rm{eff}}})_{XXZZ}} + 0.011{({{\bar k}_{\rm{eff}}})_{YYZZ}}]\\
 \times {10^{ - 7}}{\rm{Nm/}}{{\rm{m}}^{\rm{2}}}\\
{c_1} = [ - 0.085{({{\bar k}_{\rm{eff}}})_{XXXZ}} - 0.002{({{\bar k}_{\rm{eff}}})_{YYYZ}}\\
 - 0.006{({{\bar k}_{\rm{eff}}})_{ZZZX}} - 0.032{({{\bar k}_{\rm{eff}}})_{ZZZY}}\\
 - 0.002{({{\bar k}_{\rm{eff}}})_{XXYZ}} - 0.085{({{\bar k}_{\rm{eff}}})_{YYXZ}}]\\
 \times {10^{ - 7}}{\rm{Nm/}}{{\rm{m}}^{\rm{2}}}.
\end{array}
\end{equation}

\begin{figure}[!t]
\includegraphics[width=0.45\textwidth]{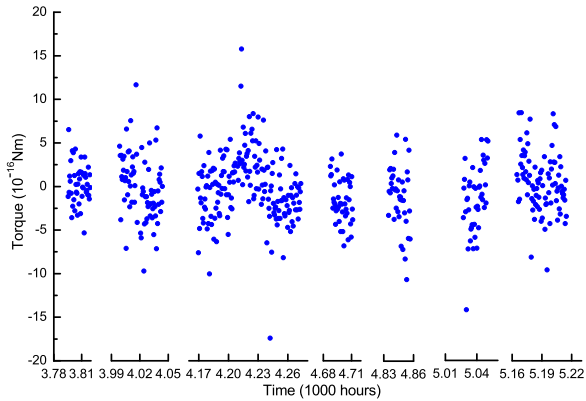}
\caption{\label{fig:fig_3} Data for $\tau _{LV}(t)$ from the HUST-2011 short-range experiment.}
\end{figure}
\begin{table} [htp]
\caption{\label{tab1:chips} Fourier amplitudes in 10$^{-16}$ Nm units. }
\begin{tabular}{|c|c|}
\hline
Mode   & Numerical value \\
\hline
$c_{0}$        &-0.22$\pm$0.95 \\
\hline
$c_{1}$      &0.13$\pm$0.22 \\
\hline
$s_{1}$        &-0.40$\pm$0.23\\
\hline
$c_{2}$      &-0.04$\pm$0.22 \\
\hline
$s_{2}$        &0.20$\pm$0.22\\
\hline
$c_{3}$      &-0.30$\pm$0.22 \\
\hline
$s_{3}$        &0.25$\pm$0.23\\
\hline
$c_{4}$      &-0.06$\pm$0.23 \\
\hline
$s_{4}$        &0.05$\pm$0.23\\
\hline
\end{tabular}
\end{table}
The systematic error of  ${\tau_{LV}}(T)$ is about 0.9$\times$10$^{-16}$ Nm, which comes from the dimensions and positions of the test and source mass,
as well as the torque calibration. The largest part of systematic error is due to the uncertainty of Newtonian torque. However, it is independent on
sidereal time $T$, and only affects the coefficient $c_{0}$. Therefore, the uncertainty of $c_{0}$ is dominated by the systematic error, and the uncertainty
of $c_{\rm{i}}$ and $s_{\rm{i}}$ (i=1,2,3,4) is dominated by the statistical error. The nine components $c_{\rm{i}}$ and $s_{\rm{i}}$ extracted from the
 experiments by Eq.(10) are listed in the Table I. Each of them is functions of the coefficients $(\bar{k}_{\rm{eff}})_{JKLM}$, such as in Eq.(11).

However, the nine signal components are insufficient to constrain independently each of the 15 degrees of freedom in $(\bar{k}_{\rm{eff}})_{JKLM}$. Following
standard practice in Ref. \cite{8}, we can obtain maximum-sensitivity constraints on each component of $(\bar{k}_{\rm{eff}})_{JKLM}$ in turn by setting the other 14 degrees of freedom to be zero. The resulting measurements and 2$\sigma$ errors on each independent component of $(\bar{k}_{\rm{eff}})_{JKLM}$ are displayed in Table II, which is in agreement with the rough estimation at level of 0.5$\times$10$^{-8}$ m$^{2}$ in Eq.(8).
 \begin{table} [!t]
\caption{\label{tab2:chips}  Coefficient values ($2\sigma$) from the experiment HUST-2011 }
\renewcommand\arraystretch{1.3}
\begin{tabular}{|c|c|}
\hline
Coefficient   & Value (10$^{-8}$m$^{2}$) \\
\hline
$(\bar{k}_{\rm{eff}})_{XXXX}$        &-0.2$\pm$2.8 \\
\hline
$(\bar{k}_{\rm{eff}})_{YYYY}$      &0.4$\pm$2.8 \\
\hline
$(\bar{k}_{\rm{eff}})_{ZZZZ}$        &-0.9$\pm$7.7\\
\hline
$(\bar{k}_{\rm{eff}})_{XXXY}$      &0.4$\pm$1.3 \\
\hline
$(\bar{k}_{\rm{eff}})_{XXXZ}$       &-0.1$\pm$0.5\\
\hline
$(\bar{k}_{\rm{eff}})_{YYYX}$     &0.6$\pm$1.3 \\
\hline
$(\bar{k}_{\rm{eff}})_{YYYZ}$       &0.4$\pm$0.5\\
\hline
$(\bar{k}_{\rm{eff}})_{ZZZX}$     &-1.3$\pm$1.4 \\
\hline
$(\bar{k}_{\rm{eff}})_{ZZZY}$        &-0.2$\pm$1.3\\
\hline
$(\bar{k}_{\rm{eff}})_{XXYY}$        &-0.1$\pm$1.7\\
\hline
$(\bar{k}_{\rm{eff}})_{XXZZ}$        &-0.2$\pm$1.0\\
\hline
$(\bar{k}_{\rm{eff}})_{YYZZ}$        &0.2$\pm$1.0\\
\hline
$(\bar{k}_{\rm{eff}})_{XXYZ}$        &0.5$\pm$0.5\\
\hline
$(\bar{k}_{\rm{eff}})_{YYXZ}$        &-0.2$\pm$0.5\\
\hline
$(\bar{k}_{\rm{eff}})_{ZZXY}$        &-0.2$\pm$0.5\\
\hline
\end{tabular}
\end{table}

The Table II represents the best measurements of possible short-range Lorentz invariance violation involving perturbative inverse-quartic corrections to Newton's law and hence of possible quadratic curvature couplings violating local Lorentz invariance. As seen above, an important feature of short-range tests of local Lorentz invariance violation in gravity is that the sensitivity to Lorentz invariance violation is dominated by edge effects. As a comparison let us consider the experiment of Indiana-2012 \cite{8}. In the experiment, each of the two masses is a planar tungsten oscillator of approximate thickness 250 $\rm{\mu}$m, separated by a gap of about 80 $\rm{\mu}$m \cite{9,10}. The force-sensitive `detector' mass is driven by the force-generating `source' mass at a resonance near 1 kHz. Detector oscillations are read out via a capacitive transducer probe coupled to a sensitive differential amplifier, with the signal fed to a lock-in amplifier referenced by the same waveform used to drive the source mass. The overlapping area of the test mass and source mass is
 $A_{p}$=5.7mm$\times$5.08mm, separated by a gap 60 $\rm{\mu}$m for $d_{\rm{min}}$ and 100 $\mu$m for $d_{\rm{max}}$. According to Eq.(7), we obtain $\Delta {C_{{\rm{Indiana}}}} \approx 15.1 \times {10^{ - 7}}{\rm{N/}}{{\rm{m}}^{\rm{2}}}$. The preliminary numerical integration gives
\begin{eqnarray}\label{equation12:eps}
\Delta F _{LV}^z \approx \!\!\!\!\!\! &&  \left[ {0.1{{({{\bar k}_{\rm{eff}}})}_{xxxx}} + 0.5{{({{\bar k}_{\rm{eff}}})}_{xxyy}} +  \cdots } \right] \nonumber\\*
&&\times {10^{ - 7}}{\rm{N/}}{{\rm{m}}^{\rm{2}}}
<  < \Delta {C_{{\rm{Indiana}}}}{\bar k_{\rm{eff}}},
\end{eqnarray}
which means the edge effect suppression factor is approximatively  $\varepsilon\sim0.03$. Therefore, the sensitivity of the apparatus to the coefficients $(\bar{k}_{\rm{eff}})_{JKLM}$ can be roughly estimated as the ratio of the thermal-noise force ($\sim$10 fN) to the scale ($\sim$0.05 $\mu$N/m$^{2}$) in Eq.(12), resulting in a level of 2$\times$10$^{-7}$ m$^{2}$.

In conclusion, in short-range tests of local Lorentz invariance violation in the gravity sector, Lorentz invariance violation force between two finite flat plates is dominated by edge effects, which adds an extra suppression factor. Nevertheless, this work presents the best current constraints of Lorentz invariance violation involving quadratic couplings of Riemann curvature in the pure gravity sector at level 10$^{-8}$ m$^{2}$.

This work was supported by the National Natural Science Foundation of China (11275075) and 111 project (B14030). MET was also funded by the Australian Research Council grant DP130100205.


\end{document}